\documentclass[a4paper]{jpconf}
\usepackage{graphicx}
\usepackage{pgfplots}
\usepackage{iopams}
\usepackage{upgreek}
\begin{document}
\title{A neural network $z$-vertex trigger for Belle II}

\author{S~Neuhaus$^1$, S~Skambraks$^2$, F~Abudinen$^3$, Y~Chen$^1$, M~Feindt$^4$, R~Fr\"uhwirth$^5$, M~Heck$^4$, C~Kiesling$^3$, A~Knoll$^1$, S~Paul$^2$ and J~Schieck$^5$}

\address{$^1$ Institute for Robotics and Embedded Systems, Technische Universit\"at M\"unchen, Germany}
\address{$^2$ Physik Department, Technische Universit\"at M\"unchen, Germany}
\address{$^3$ Max-Planck-Institut f\"ur Physik, M\"unchen, Germany}
\address{$^4$ Institute for Experimental Nuclear Physics, Karlsruher Institut f\"ur Technologie, Germany}
\address{$^5$ Institute for High Energy Physics, \"Osterreichische Akademie der Wissenschaften, Austria}

\ead{sara.neuhaus@in.tum.de}

\begin{abstract}
We present the concept of a track trigger for the Belle~II experiment, based on a neural network approach, that is able to reconstruct the $z$ (longitudinal) position of the event vertex within the latency of the first level trigger. The trigger will thus be able to suppress a large fraction of the dominating background from events outside of the interaction region. The trigger uses the drift time information of the hits from the Central Drift Chamber (CDC) of Belle~II within narrow cones in polar and azimuthal angle as well as in transverse momentum (sectors), and estimates the $z$-vertex without explicit track reconstruction. The preprocessing for the track trigger is based on the track information provided by the standard CDC trigger. It takes input from the 2D ($r-\varphi$) track finder, adds information from the stereo wires of the CDC, and finds the appropriate sectors in the CDC for each track in a given event. Within each sector, the $z$-vertex of the associated track is estimated by a specialized neural network, with a continuous output corresponding to the scaled $z$-vertex. The input values for the neural network are calculated from the wire hits of the CDC.
\end{abstract}

\section{Introduction to the Belle~II track trigger}
The Belle~II experiment~\cite{Belle2TDR} will go into operation at the upgraded KEKB collider (SuperKEKB~\cite{superkekb2}) in 2016. SuperKEKB is designed to deliver an instantaneous luminosity $\mathcal{L}=8\cdot10^{35}\,\mathrm{cm}^{-2}\mathrm{s}^{-1}$, a factor of 40 larger than the previous KEKB world record. The Belle~II experiment will therefore have to cope with a much larger machine background than its predecessor Belle, dominated by Touschek scattering~\cite{Touschek,Piwinski}. This background produces a high rate of background events with the longitudinal position of the vertex ($z$-vertex) outside of the nominal interaction region, where the physically interesting reactions from the $e^+e^-$ collisions are produced. By reconstructing the $z$-vertex within the latency of the first level (L1) trigger, a large fraction of this background can be suppressed. The total latency of $5\,\upmu\mathrm{s}$ for the full L1 trigger allows for only $\approx 1\,\upmu\mathrm{s}$ for the $z$-vertex reconstruction. Neural networks of the Multi Layer Perceptron (MLP) type are well suited to meet these constraints due to their inherent parallelism and deterministic runtime. The approach presented here uses MLPs with only the hits in the Central Drift Chamber (CDC) as input to make an estimation of the $z$-vertex without explicit track reconstruction, based on the concepts presented in~\cite{SSMA, FAMA, RT2014}.

The CDC contains 56 layers of sense wires of either ``axial'' orientation (aligned with the solenoidal magnetic field of 1.5\,T along the $z$-axis) or ``stereo'' (skewed with respect to the axial wires). 6 or 8 adjacent layers have the same orientation and number of wires per layer and are combined in a so-called SuperLayer~(SL). In total there are 9~SL alternating between axial and stereo SL. By combining the information of axial and stereo wires it is possible to reconstruct a full 3D helix track, which can be parametrized by $(p_T, \varphi, \theta, z, d)$, where $p_T$ is the transverse momentum, $\varphi$ and $\theta$ are the azimuthal and polar angle at the vertex, $z$ is the vertex position along the beamline and $d$ is the displacement of the vertex from the beamline in the $r-\varphi$ plane. The latter is expected to be small and is therefore assumed to be $d = 0\,\mathrm{cm}$ in our studies.

\label{sec:cdctrigger}
\begin{figure}[t]
\centering
\begin{minipage}[b]{0.73\textwidth}
\begin{tikzpicture}
	\tikzstyle{every node}=[font=\small]
	\def\xcdc{-2};
	\def\xtsf{0};
	\def\xtime{2.5};
	\def\xnn{5};
	\def\xgdl{7};
	\node[draw,rounded corners,fill=black!10,align=center] (cdc) at (\xcdc,1) {CDC\\Readout};
	\node[draw,rounded corners,fill=black!10,align=center] (axial) at (\xtsf,2) {Axial\\TSF};
	\node[draw,rounded corners,fill=black!10,align=center] (stereo) at (\xtsf,0) {Stereo\\TSF};
	\node[draw,rounded corners,fill=black!10,align=center] (2D) at (\xtime,2) {2D\\Trigger};
	\node[draw,rounded corners,fill=black!10,align=center] (3D) at (\xnn,2) {3D\\Trigger};
	\node[draw,rounded corners,fill=black!30,align=center] (NN) at (\xnn,0) {Neural\\Trigger};
	\node[draw,rounded corners,fill=black!10,align=center] (time) at (\xtime,0) {Event\\Time};
	\node[draw,rounded corners,fill=black!10,align=center] (gdl) at (\xgdl,1) {Global\\Decision\\Logic};
	\draw[-latex] (cdc) |- (axial);
	\draw[-latex] (cdc) |- (stereo);
	\draw[-latex,very thick] (axial) -- node[sloped,above] {$\mathrm{TS}_\mathrm{axial}$} (2D);
	\coordinate (axialtime) at (\xtime/2,0.2);
	\draw[-latex] (axial.east) |- (\xtime/2,1.8) -- (axialtime) -- (axialtime -| time.west);
	\draw[-latex] (stereo) -- (time);
	\draw[-latex] (stereo) -- (\xtsf,1.3) -| (3D);
	\draw[-latex,very thick] (stereo) -- (\xtsf,-0.85) -| node[sloped,above,pos=0.38] {$\mathrm{TS}_\mathrm{stereo}$} (NN);
	\draw[-latex] (2D) -- (3D);
	\draw[-latex,very thick] (2D) -- (\xtime,1.1) -| node[sloped,below,near start] {$\mathrm{TS}_\mathrm{axial}$, $p_T$, $\varphi$} (NN);
	\coordinate (2Dgdl) at (\xgdl+0.1,2.75);
	\draw[-latex] (2D) |- (2Dgdl) -- (2Dgdl |- gdl.north);
	\coordinate (time3D) at (\xtime/2+\xnn/2,1.8);
	\draw[-latex] (time.east) |- (\xtime/2+\xnn/2,0.2) -- (time3D) -- (time3D -| 3D.west);
	\draw[-latex] (time) -- (NN);
	\coordinate (timegdl) at (\xgdl+0.1,-1.1);
	\draw[-latex] (time) |- (timegdl) -- (timegdl |- gdl.south);
	\coordinate (3Dgdl) at (\xgdl-0.1,2);
	\draw[-latex] (3D) -- (3Dgdl) -- (3Dgdl |- gdl.north);
	\coordinate (NNgdl) at (\xgdl-0.1,0);
	\draw[-latex,very thick] (NN) -- node[sloped,below] {$z$} (NNgdl) -- (NNgdl |- gdl.south);
\end{tikzpicture}
\end{minipage}
\begin{minipage}[b]{0.25\textwidth}
\caption{Schematic of the CDC trigger. The proposed neural $z$-vertex trigger takes input from the TS finders and from the 2D trigger.}
\end{minipage}
\label{fig:signalflow}
\end{figure}
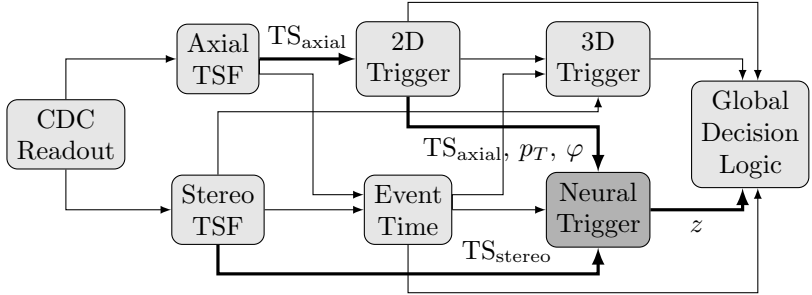

A schematic view of the track trigger is shown in figure~\ref{fig:signalflow}. For each of the 9~SL the wire hits are combined in so-called Track Segments~(TS) by the Track Segment Finder (TSF) to reduce noise and the amount of data for the trigger. A TS is a predefined arrangement of several wires from 5~layers within a SL which produces a hit if at least 4~wires in different layers are active. For each TS the TSF transmits an id, the drift time of a reference wire in the TS with a resolution of 2\,ns and the position of the track relative to this wire. The latter is obtained from a lookup table based on the hit pattern within the TS and can be \emph{left}, \emph{right} or \emph{undecided}\footnote{The left/right information was not included in earlier studies \cite{SSMA, FAMA, RT2014}, which were based on an older version of the Belle~II simulation.}. The event time that is needed to calculate the absolute drift times is estimated from the fastest wire hits in an event.

The 2D trigger combines TS from the axial SL to provide 2D tracks in $r-\varphi$ space~\cite{iwasaki}. The 3D trigger then adds the information from the stereo TS to determine the $z$-vertex. The proposed neural trigger will run in parallel to the conventional analytical 3D trigger and employ MLPs trained on sample tracks to predict the $z$-vertex. The outputs of the 2D/3D triggers are fed into the Global
Decision Logic, which combines the results of all subtriggers
and makes the final decision. More details about the Belle~II detector and the trigger system can be found in~\cite{Belle2TDR, iwasaki}.

\section{Structure of the neural network $z$-vertex trigger}
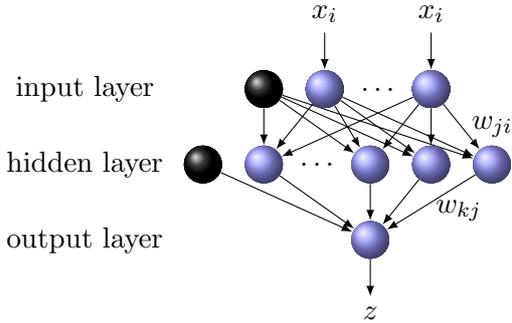
\begin{figure}[t]
\begin{minipage}[b]{0.48\textwidth}
\begin{tikzpicture}
  \def\yin{1};
  \def\yhid{0};
  \def\yout{-1};
  \node[ball color=blue!40,circle,minimum size=0.5cm] (out) at (0,\yout) {};
  \draw[-latex] (out) -- (0,\yout-0.75) node[below] {$z$};
  \node (morehidden) at (-0.7,\yhid) {$\dots$};
  \node (moreinput) at (0.1,\yin) {$\dots$};
  \node[ball color=black!100,circle,minimum size=0.5cm] (biashidden) at (-2.2,\yhid) {};
  \node[ball color=black!100,circle,minimum size=0.5cm] (biasinput) at (-1.4,\yin) {};
  \draw[-latex] (biashidden) -- (out);
  \node[ball color=blue!40,circle,minimum size=0.5cm] (hidden) at (1.6,\yhid) {};
  \draw[-latex] (hidden) -- node[right,pos=0.65,xshift=3pt]{$w_{kj}$} (out);
  \draw[-latex] (biasinput) -- (hidden);
  \node[circle,minimum size=0.5cm] (in) at (0.8,\yin) {};
  \draw[-latex] (in) -- node[right]{$w_{ji}$} (hidden);
  \node[circle,minimum size=0.5cm] (in) at (-0.6,\yin) {};
  \draw[-latex] (in) -- (hidden);
  \foreach \inputnode in {0.8,-0.6}
  { \node[circle,minimum size=0.5cm] (in) at (\inputnode,\yin) {};
	\draw[latex-] (in) -- (\inputnode,\yin+0.75) node[above] {$x_i$}; }
  \foreach \hidden in {0.8,0,-1.4}
  { \node[ball color=blue!40,circle,minimum size=0.5cm] (hidden) at (\hidden,\yhid) {};
    \draw[-latex] (hidden) -- (out);
    \draw[-latex] (biasinput) -- (hidden);
    \foreach \inputnode in {0.8,-0.6}
    { \node[ball color=blue!40,circle,minimum size=0.5cm] (in) at (\inputnode,\yin) {};
      \draw[-latex] (in) -- (hidden); } }
  \draw (-3.75,\yin) node[align=center] {input layer};
  \draw (-3.75,\yhid) node[align=center] {hidden layer};
  \draw (-3.75,\yout) node[align=center] {output layer};
\end{tikzpicture}
\end{minipage}
\begin{minipage}[b]{0.5\textwidth}
\caption{Structure of an MLP with one hidden layer. In the nodes the weighted sum of the input values $x_i$ is evaluated by the activation function to produce one output value. The black nodes denote constant bias nodes. Input and target values are always scaled to the co-domain of the activation function $[-1,1]$.}
\end{minipage}
\label{fig:mlp}
\end{figure}

The core concept of the neural trigger are networks of the Multi Layer Perceptron (MLP) type. The MLP is a universal function approximator~\cite{perceptron-hypothesis} with the structure of an acyclic graph (see figure~\ref{fig:mlp}). Each neuron calculates a weighted sum of the input values and evaluates it with a non-linear bounded activation function, in our case the $\tanh$ function:
\begin{equation}
  y_j = \tanh\left( \sum_{i=0}^N{w_{ij} \cdot x_i} \right)
\end{equation}
where $x_i$ are the input values and $w_{ij}$ are the weights, which contain the information of the calculated function. The first input is $x_0 = 1$ such that the corresponding weight $w_{j0}$ acts as a constant bias. During training the weights of the MLP are adjusted with a back propagation algorithm such that for given input values the output values converge to certain target values. These target values are the $z$-vertex or other track parameters, scaled to the co-domain of the activation function $[-1,1]$. The input values are calculated from the TS hits and scaled to the same range, following different input representations that are explained in detail below.

We use an ensemble of ``expert'' MLPs that are specialized to small sectors in phase space~\cite{SSMA}. Only tracks with helix parameters from limited intervals are selected as training samples for a given sector. For each sector a number of ``relevant'' TS are identified based on a histogram of active TS generated from tracks within the sector. Only hits from the relevant TS are considered for the MLP, limiting the amount of input data for each sector. While the sector in $p_T$ and $\varphi$ can be selected using the estimation of the 2D trigger, the polar angle $\theta$ is unknown. As the stereo hits depend on both $z$ and $\theta$, one of them cannot be reconstructed precisely without knowing the other. Therefore we use a multi-step cascade prediction~\cite{SSMA, RT2014}, where a primary MLP specialized to a small sector in $p_T$ and $\varphi$ is used to predict $z$ and $\theta$. This prediction is used to select another MLP restricted in all track parameters, which can in the ideal case predict $z$ with the desired resolution. Otherwise the concept can be extended by adding another step, as was successfully demonstrated in \cite{RT2014}.

\begin{figure}[b]
\begin{minipage}[b]{0.51\textwidth}
\begin{tikzpicture}[scale=0.9]
  \draw[gray] (185:3.4641) arc (185:-5:3.4641);
  \draw[gray] (0,0) -- (95:3.4641);
  \draw[-latex] (-4,0) -- (4,0);
  \draw[-latex] (0,-0.5) -- (0,4);
  \draw (-3.4641,2pt) -- (-3.4641,-2pt) node[below] {$-r$};
  \draw (3.4641,2pt) -- (3.4641,-2pt) node[below] {$r$};
  \draw (-2pt,3.4641) -- (2pt,3.4641) node[right] {$r$};
  \draw (133:2) -- (137:2) node[xshift=-5pt,yshift=-5pt] {$R$};
  \draw (134:4) -- (136:4) node[left] {$2R$};
  \draw[-latex,thick] (0,0) arc (-45:175:2);
  \draw[dashed] (0,0) --  (105:3.4641) -- (135:4) -- cycle;
  \fill (95:3.4641) circle(2pt);
  \draw[dashed] (0,0) -- (3,3);
  \draw (105:2.5) arc (105:95:2.5) node[below,xshift=2pt] {$\varphi_\mathrm{rel}$};
  \draw (95:1) arc (95:0:1) node[xshift=-13pt,yshift=9pt] {$\varphi_\mathrm{wire}$};
  \draw (45:1.6) arc (45:0:1.6) node[xshift=-12pt,yshift=15pt] {$\varphi$};
  \draw (105:1.9) arc (105:45:1.9) node[xshift=-20pt,yshift=2pt] {$\varphi_{p_T}$};
\end{tikzpicture}
\end{minipage}
\begin{minipage}[b]{0.47\textwidth}
\caption{Using the estimation for the curvature and the azimuthal angle provided by the 2D track finder, the relative azimuthal angle between the track and a given wire can be calculated.}
\end{minipage}
\label{fig:2dcorrection}
\end{figure}
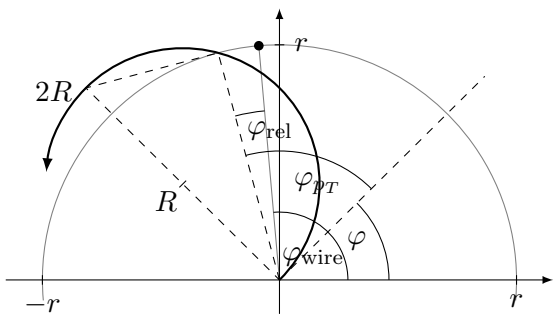

To find an input representation for the MLP of a given sector, the relevant TS must be mapped to the input nodes. The most straightforward way to do this is to use one input node per relevant TS, with the drift time scaled to $[-1,1]$ as input value and the left/right information encoded in the sign (``rep1''). For undecided left/right information the input value is $0$, which is also the default value for inactive TS. While this mapping produces good results for small sectors, it is not an efficient representation for larger sectors, as the number of relevant TS grows with the sector size while the typical number of hits per track is just one hit per SL. In a second approach we therefore fix the number of inputs by using always one hit per SL, choosing the hit with the shortest drift time in case of several neighboring hits. In addition to the drift time we also use the wire position as input for the MLP, conveniently encoded in the TS id within a SL, which is proportional to a discrete azimuthal angle. This input representation (``rep2'') results in a constant number of 18 inputs, independent of the sector size.

Making use of the $\varphi$ and $p_T$ estimation of the 2D trigger it is possible to reduce the number of sectors by calculating the azimuthal angle \emph{relative} to the reconstructed 2D track and use this as input for the MLP together with the drift times (``rep3''). The input representation is then independent of $\varphi$ and a single sector\footnote{The hardware implementation for a fully pipe-lined neural trigger will consist of 4~FPGA boards, each covering a quarter of the detector. This hardware sectorization is used also by the 2D trigger to cope with the high bandwidth requirements.} is enough for this variable. The relative wire position can be obtained from the position of the wire and the estimated 2D values (see figure~\ref{fig:2dcorrection}):
\begin{equation}
  \varphi_\mathrm{rel} = \varphi_\mathrm{wire} - \varphi - \arcsin\left( \frac{r}{2R} \right) \equiv \varphi_\mathrm{wire} - \varphi - \varphi_{p_T}
\end{equation}
where $\varphi$ is the estimated track direction at the vertex, $r$ is the radial position of the wire, $R \propto p_T$ is the radius of the helix and $\varphi_\mathrm{wire}$ is the azimuthal coordinate of a reference position on the wire. $\varphi_{p_T} = \arcsin\left( \frac{r}{2R} \right)$ is the change in the azimuthal angle due to the curling of the track in the magnetic field and is also proportional to the arc length $\mu$ of the track in the $r$-$\varphi$-plane:
\begin{equation}
  \mu(r, R) = 2 \varphi_{p_T} R
\end{equation}
In the $\mu$-$z$-projection the track is linear, facilitating the extrapolation of the track to the vertex:
\begin{equation}
  z(r) = \mu(r,R) \cot(\theta) + z_0
\end{equation}
where $z_0$ is the $z$-coordinate of the vertex and $\theta$ is the polar angle at the vertex. Therefore the scaled arc length $\mu(r,R)$ for each SL is used as an additional input for the MLP, leading to a total of 27 inputs (``rep4'').

\section{Experimental comparison of input representations}
To evaluate the different input representations we have used simulated single muon tracks generated within limited ranges of $p_T$, $\varphi$, $\theta$ and $z$. The generated distribution within these ranges is uniform, except for $p_T$ where we use a uniform distribution in $p_T^{-1}$, i.e. in the curvature. When sectorizing $p_T$ we also choose bins of equal size in $p_T^{-1}$. Ranges in $\varphi$ apply only to the simpler input representations rep1 and rep2, otherwise we use a whole quarter of the detector ($\varphi \in [135^\circ,225^\circ]$). The Geant4 based detector simulation of the Belle~II CDC includes material interaction with the inner detector components (beampipe and vertex detectors), non-linear drift velocity due to inhomogeneities in the electric field and the gravitational wire sag effect.

\begin{table}[t]
\caption{Estimates for the precision of the $z$ prediction in small sectors for different input representations. $\Delta z$ is the RMS of the difference between true and predicted $z$-vertex.}
\begin{center}
\begin{tabular}{llllll}
\br
& input & high $p_T$ & & low $p_T$\\
& values & $\Delta z$ & inputs & $\Delta z$ & inputs\\
\mr
rep1 & 1 per TS $(t)$& 1.40\,cm & 17 & 1.89\,cm & 31\\
rep2 & 2 per SL $(t, \varphi_\mathrm{wire})$ & 1.10\,cm & 18 & 1.50\,cm & 18\\
rep3 & 2 per SL $(t, \varphi_\mathrm{rel})$ & 1.18\,cm & 18 & 1.32\,cm & 18\\
rep4 & 3 per SL $(t, \varphi_\mathrm{rel}, \mu)$ & 1.15\,cm & 27 & 1.18\,cm & 27\\
\br
\end{tabular}
\end{center}
\label{tab:perfect}
\end{table}

The presented input representations are first studied with very small sectors to compare their performance under optimal conditions. We train one sector for low $p_T$ tracks (${p_T^{-1} \in [1.95,2]\,\mathrm{GeV}^{-1}}$) and one for high $p_T$ tracks ($p_T^{-1} \in [0.2,0.25]\,\mathrm{GeV}^{-1}$). The remaining track parameters are constrained to $\varphi \in [180^\circ,181^\circ]$, $\theta \in [50^\circ,60^\circ]$ and $z \in [-10,10]\,\mathrm{cm}$. The results are summarized in table~\ref{tab:perfect}. Based on the $z$-vertex distribution in Belle~\cite{Belle2TDR} we estimate that a resolution of $\approx 2\,\mathrm{cm}$ will be required to efficiently suppress the background, which is reached in both tested sectors for all input representations. Fixing the network size (rep2, rep3 and rep4) shows an advantage not only for the hardware implementation but also for the resolution. For the low $p_T$ case the resolution improves when correcting the curvature (rep3) and even more so when explicitly using the arc length $\mu$ (rep4).

\begin{figure}[t]
\begin{center}
\begin{tikzpicture}[scale=0.66]
\draw[-latex] (0,0) -- node[yshift=-0.75cm] {$p_T^{-1}\,[\mathrm{GeV}^{-1}]$} (11,0);
\draw (1.16,1pt) -- (1.16,-1pt) node[below] {0.5};
\draw[very thin,gray] (1.16,0) -- (1.16,5.0);
\draw (3.09,1pt) -- (3.09,-1pt) node[below] {1};
\draw[very thin,gray] (3.09,0) -- (3.09,5.0);
\draw (5.01,1pt) -- (5.01,-1pt) node[below] {1.5};
\draw[very thin,gray] (5.01,0) -- (5.01,5.0);
\draw (6.94,1pt) -- (6.94,-1pt) node[below] {2};
\draw[very thin,gray] (6.94,0) -- (6.94,5.0);
\draw (8.87,1pt) -- (8.87,-1pt) node[below] {2.5};
\draw[very thin,gray] (8.87,0) -- (8.87,5.0);
\draw[-latex] (0,0) -- (0,5.0) node[above] {$\Delta \theta$};
\draw (1pt,0.5) -- (-1pt,0.5) node[left] {$1^\circ$};
\draw[very thin,gray] (0,0.5) -- (11,0.5);
\draw (1pt,1.0) -- (-1pt,1.0) node[left] {$2^\circ$};
\draw[very thin,gray] (0,1.0) -- (11,1.0);
\draw (1pt,1.5) -- (-1pt,1.5) node[left] {$3^\circ$};
\draw[very thin,gray] (0,1.5) -- (11,1.5);
\draw (1pt,2.0) -- (-1pt,2.0) node[left] {$4^\circ$};
\draw[very thin,gray] (0,2.0) -- (11,2.0);
\draw (1pt,2.5) -- (-1pt,2.5) node[left] {$5^\circ$};
\draw[very thin,gray] (0,2.5) -- (11,2.5);
\draw (1pt,3.0) -- (-1pt,3.0) node[left] {$6^\circ$};
\draw[very thin,gray] (0,3.0) -- (11,3.0);
\draw (1pt,3.5) -- (-1pt,3.5) node[left] {$7^\circ$};
\draw[very thin,gray] (0,3.5) -- (11,3.5);
\draw (1pt,4.0) -- (-1pt,4.0) node[left] {$8^\circ$};
\draw[very thin,gray] (0,4.0) -- (11,4.0);
\draw (1pt,4.5) -- (-1pt,4.5) node[left] {$9^\circ$};
\draw[very thin,gray] (0,4.5) -- (11,4.5);
\tikzset{every path/.style={black,dotted,very thick}}
\draw (0.2,1.15) -- (0.6,1.18) -- (1.0,1.28) -- (1.4,1.09) -- (1.8,1.25) -- (2.2,1.45) -- (2.6,1.36) -- (3.0,1.27) -- (3.4,1.33) -- (3.8,1.62) -- (4.2,1.97) -- (4.6,1.59) -- (5.0,1.78) -- (5.4,1.55) -- (5.8,1.99) -- (6.2,1.97) -- (6.6,1.85) -- (7.0,2.20) -- (7.4,2.09) -- (7.8,2.43) -- (8.2,2.37) -- (8.6,2.19) -- (9.0,2.55) -- (9.4,2.59) -- (9.8,2.81) -- (10.2,3.13) -- (10.6,3.21);
\tikzset{every path/.style={red,dashed,thick}}
\draw (0.2,1.73) -- (0.6,1.45) -- (1.0,1.52) -- (1.4,1.50) -- (1.8,1.62) -- (2.2,1.63) -- (2.6,1.78) -- (3.0,1.88) -- (3.4,2.34) -- (3.8,1.77) -- (4.2,1.96) -- (4.6,2.02) -- (5.0,1.91) -- (5.4,2.48) -- (5.8,2.37) -- (6.2,2.34) -- (6.6,2.63) -- (7.0,2.50) -- (7.4,2.27) -- (7.8,2.70) -- (8.2,2.69) -- (8.6,3.34) -- (9.0,3.04) -- (9.4,3.41) -- (9.8,3.59) -- (10.2,3.97) -- (10.6,4.17);
\tikzset{every path/.style={blue}}
\draw (0.2,1.92) -- (0.6,1.62) -- (1.0,1.60) -- (1.4,1.77) -- (1.8,2.00) -- (2.2,1.94) -- (2.6,1.79) -- (3.0,1.73) -- (3.4,2.05) -- (3.8,2.16) -- (4.2,2.40) -- (4.6,2.38) -- (5.0,2.13) -- (5.4,2.32) -- (5.8,2.67) -- (6.2,2.69) -- (6.6,2.76) -- (7.0,2.92) -- (7.4,2.81) -- (7.8,3.06) -- (8.2,3.54) -- (8.6,3.41) -- (9.0,3.63) -- (9.4,3.47) -- (9.8,3.68) -- (10.2,3.61) -- (10.6,3.67);
\end{tikzpicture}
\begin{tikzpicture}[scale=0.66]
\draw[-latex] (0,0) -- node[yshift=-0.75cm] {$p_T^{-1}\,[\mathrm{GeV}^{-1}]$} (11,0);
\draw (1.16,1pt) -- (1.16,-1pt) node[below] {0.5};
\draw[very thin,gray] (1.16,0) -- (1.16,5.0);
\draw (3.09,1pt) -- (3.09,-1pt) node[below] {1};
\draw[very thin,gray] (3.09,0) -- (3.09,5.0);
\draw (5.01,1pt) -- (5.01,-1pt) node[below] {1.5};
\draw[very thin,gray] (5.01,0) -- (5.01,5.0);
\draw (6.94,1pt) -- (6.94,-1pt) node[below] {2};
\draw[very thin,gray] (6.94,0) -- (6.94,5.0);
\draw (8.87,1pt) -- (8.87,-1pt) node[below] {2.5};
\draw[very thin,gray] (8.87,0) -- (8.87,5.0);
\draw[-latex] (0,0) -- (0,5.0) node[above,xshift=2] {$\Delta z\,[\mathrm{cm}]$};
\draw (1pt,0.5) -- (-1pt,0.5) node[left] {1};
\draw[very thin,gray] (0,0.5) -- (11,0.5);
\draw (1pt,1.0) -- (-1pt,1.0) node[left] {2};
\draw[very thin,gray] (0,1.0) -- (11,1.0);
\draw (1pt,1.5) -- (-1pt,1.5) node[left] {3};
\draw[very thin,gray] (0,1.5) -- (11,1.5);
\draw (1pt,2.0) -- (-1pt,2.0) node[left] {4};
\draw[very thin,gray] (0,2.0) -- (11,2.0);
\draw (1pt,2.5) -- (-1pt,2.5) node[left] {5};
\draw[very thin,gray] (0,2.5) -- (11,2.5);
\draw (1pt,3.0) -- (-1pt,3.0) node[left] {6};
\draw[very thin,gray] (0,3.0) -- (11,3.0);
\draw (1pt,3.5) -- (-1pt,3.5) node[left] {7};
\draw[very thin,gray] (0,3.5) -- (11,3.5);
\draw (1pt,4.0) -- (-1pt,4.0) node[left] {8};
\draw[very thin,gray] (0,4.0) -- (11,4.0);
\draw (1pt,4.5) -- (-1pt,4.5) node[left] {9};
\draw[very thin,gray] (0,4.5) -- (11,4.5);
\tikzset{every path/.style={black,dotted,very thick}}
\draw (0.2,1.57) -- (0.6,1.44) -- (1.0,1.66) -- (1.4,1.36) -- (1.8,1.79) -- (2.2,1.55) -- (2.6,1.71) -- (3.0,1.56) -- (3.4,1.64) -- (3.8,1.74) -- (4.2,1.65) -- (4.6,1.78) -- (5.0,2.25) -- (5.4,1.93) -- (5.8,2.15) -- (6.2,2.03) -- (6.6,2.50) -- (7.0,2.74) -- (7.4,2.00) -- (7.8,3.28) -- (8.2,2.70) -- (8.6,2.20) -- (9.0,2.67) -- (9.4,3.05) -- (9.8,2.79) -- (10.2,3.42) -- (10.6,3.29);
\tikzset{every path/.style={red,dashed,thick}}
\draw (0.2,1.92) -- (0.6,1.76) -- (1.0,2.23) -- (1.4,2.11) -- (1.8,2.15) -- (2.2,1.81) -- (2.6,2.07) -- (3.0,1.96) -- (3.4,2.48) -- (3.8,1.85) -- (4.2,2.21) -- (4.6,2.13) -- (5.0,2.26) -- (5.4,2.08) -- (5.8,2.83) -- (6.2,2.64) -- (6.6,2.77) -- (7.0,2.83) -- (7.4,2.81) -- (7.8,2.93) -- (8.2,3.13) -- (8.6,3.24) -- (9.0,3.45) -- (9.4,3.76) -- (9.8,3.57) -- (10.2,4.25) -- (10.6,4.15);
\tikzset{every path/.style={blue}}
\draw (0.2,2.25) -- (0.6,1.94) -- (1.0,1.81) -- (1.4,1.96) -- (1.8,2.11) -- (2.2,2.71) -- (2.6,2.15) -- (3.0,2.12) -- (3.4,2.25) -- (3.8,2.11) -- (4.2,2.33) -- (4.6,2.37) -- (5.0,2.81) -- (5.4,2.33) -- (5.8,3.00) -- (6.2,3.08) -- (6.6,3.06) -- (7.0,3.33) -- (7.4,3.10) -- (7.8,3.42) -- (8.2,4.07) -- (8.6,3.25) -- (9.0,3.73) -- (9.4,3.61) -- (9.8,3.85) -- (10.2,3.76) -- (10.6,3.82);
\end{tikzpicture}
\end{center}
\caption{$\theta$ and $z$ prediction without a priori 3D knowledge for different input representations depending on $p_T^{-1}$. Blue solid line: rep2, red dashed line: rep3, black dotted line: rep4.}
\label{fig:primary}
\end{figure}
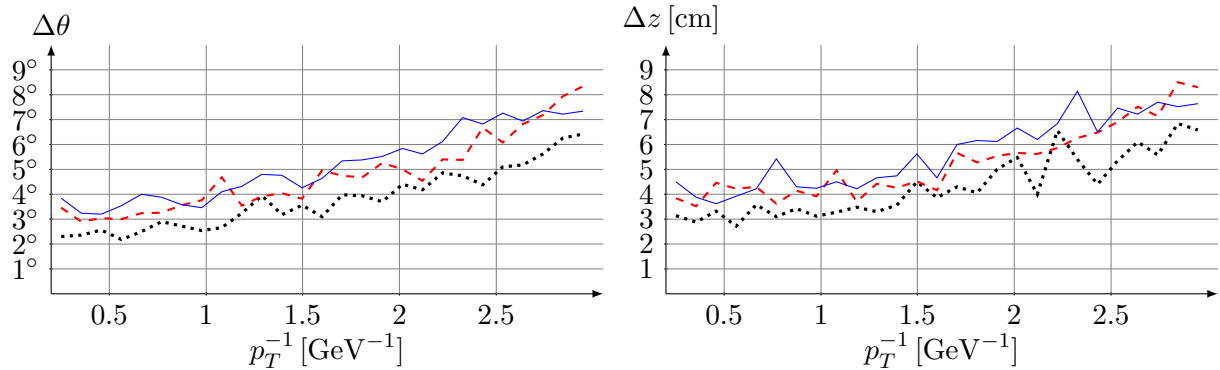

Next we compare the prediction capability for the first step of the cascade prediction, when no information about $\theta$ and $z$ is available. In $\varphi$ and $p_T^{-1}$ we use a gaussian smearing with $\sigma_\varphi = 1^\circ$, $\sigma_{p_T^{-1}} = 0.1\,\mathrm{GeV}^{-1}$ on the simulated values before using them to simulate the effect of errors in the 2D trigger prediction. The 2D trigger is still under construction and not yet fully tested, but will likely exceed this resolution. The track parameters are constrained to $p_T^{-1} \in [0.2,3]\,\mathrm{GeV}^{-1}$, $\varphi \in [179^\circ,181^\circ]$, $\theta \in [35^\circ,123^\circ]$ and $z \in [-50,50]\,\mathrm{cm}$. The $p_T$ range is divided into 9 sectors of width $\Delta p_T^{-1} = 0.31\,\mathrm{GeV}^{-1}$. In figure~\ref{fig:primary} the resolution is shown depending on $p_T^{-1}$. rep1 is not included in the test, as it was least effective in the previous test and the MLPs are too big to train efficiently. The uncertainty of both the $z$ prediction and the $\theta$ prediction increases with rising curvature for all tested input representations. The best results are again achieved for rep4. Although the target resolution is not reached, the prediction of the primary MLP can be used to significantly decrease the sector size for the next step.

Even in this first step the MLP prediction exceeds the estimate for the 3D trigger of $\Delta z = 9.3\,\mathrm{cm}$ given in \cite{Belle2TDR}, where no drift times were used. Recent progress in the development of the 3D trigger including the drift time information suggests that a better $z$ resolution can be achieved. A comparison of the neural trigger and the 3D trigger will be subject of future studies.

\section{Conclusion}
We have presented a setup for a first level $z$-vertex trigger based on an ensemble of ``expert'' MLPs, specialized to sectors in phase space, and demonstrated the influence of different input representations on the final $z$-vertex resolution. Efficiently using the information provided by earlier trigger components and knowledge about the detector geometry itself, we can reduce not only the size of each expert network, but also the total number of sectors.

In future investigations we will try to improve the resolution for low momentum tracks, combining different training algorithms for the neural networks with the presented preprocessing. The fully pipe-lined neural network trigger will be implemented in hardware on Virtex~7 FPGA boards and installed in the Belle~II detector.

\section*{References}
\bibliographystyle{iopart-num}
\bibliography{references}

\end{document}